\documentclass[aip,reprint,amsmath,amssymb,reprint,twocolumn
]{revtex4-1}

\usepackage{graphicx}
\usepackage{dcolumn}
\usepackage{bm}
\usepackage{hyperref}
\hypersetup{
    colorlinks,
    citecolor=blue,
    filecolor=black,
    linkcolor=black,
    urlcolor=blue
}

\begin{document}


\title{Carrier transport in layered nanolaminated films }

\author{Aniruddha Konar}
\email{aniruddha.konar@globalfoundries.com}
\affiliation{GLOBALFOUNDRIES, Bangalore, 560045, India. }
\author{ Rajan K. Pandey}
\affiliation{GLOBALFOUNDRIES, Bangalore, 560045, India. }
\author{ Tamilmani Ethirajan}
\affiliation{GLOBALFOUNDRIES, Bangalore, 560045, India. }

\date{\today}

\begin{abstract}   Analyzing {\it{ab-initio}} electronic and phonon band structure, temperature-dependent  carrier transport  in
layered Ti$_{2}$AlC is investigated.  It is found that cylindrical Fermi surface is the origin of the anisotropic carrier effective mass
(infinite effective mass along $c$ axis ) leading to strong anisotropic (insulator along $c$ axis and metallic along the layer) carrier transport in these films. 
Using electronic and phonon bandstructures, we develop an analytical model of electron-phonon interaction as well as in-plane carrier conductivity originating from strong inter-valley (s$\rightarrow$d) scattering in Ti$_{2}$AlC.
We   invoke  density   functional  theory  to   calculate  the deformation  potential   corresponding  to  acoustic  phonon vibration.  The
calculated  deformation  potential  is  in  well  agreement  with  the
extracted deformation potential value from the transport data. Extracted deformation potential will be useful for prediction of transport quantities for application of these metals at elevated temperatures.
\end{abstract}
\maketitle
\section{introduction}
Layered nano-laminated ternary carbide and nitride metals known as MAX phases, are emerging new class of materials (metals) that bridge the gap of properties between metals and ceramics\cite{Radovic_ACSB13}. 
Commonly denoted   as M$_{n+1}$AX$_{n}$,   these materials are known for their unusual and exotic   metallic    and   ceramic
properties \cite{Book_Barsoum13,Sun_Review11,Book_Barsoum00}. Here  $M$ is an transition  metal (M= Ti,  Cr, V, etc)
and A  is usually an element  from group IIIA  or IV, and X  is either
carbon or nitrogen.    
These are potential compounds for many emerging applications  due  to their  combined
metallic and ceramic properties  such as high thermal \cite{Book_Barsoum13}, and electrical conductivity \cite{FinklePRB03,HettingerPRB05}, high stiffness \cite{RadovicActa02,BarsoumNature03} ,
oxidation  resistance  and  high  damage  tolerance and low thermal expansion coefficients \cite{Book_Barsoum13}.  Due  to  their  high
conductivity and  low friction, these  compounds are best  suitable to
rotating electrical  contacts and high  temperature applications.  Recently, analogous to two-dimensional (2D) semiconductors \cite{NovoselovNature07,WangNature12} , 2D transparent metallic layers known as MXenes \cite{HalimCM14,GhidiuNature14} are realized by chemically removing $A$ species from single layer of MAX compounds. These 2D metallic layers can be suitable materials for electrical contacts in emerging 2D electronics \cite{NovoselovNature07, KonarNaturecomm12} as well as transparent electrodes in photovoltaic devices. 
\begin{figure}[t]
\includegraphics[width=88 mm]{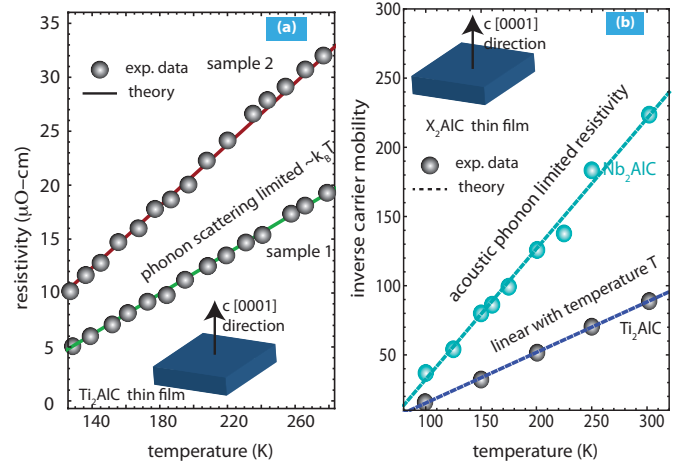}
\caption{(a) carrier resistivity function of temperature
(data taken from  Ref.\cite{MauchampPRB13}) for two different $X_{2}$AlC (X=Ti) metal films,
(b) measured inverse-mobility ($\approx$ resistivity)  data  with
temperatures of $X_{2}$AlC (X=Ti, Nb, etc) metal films
(taken from Ref.\cite{HettingerPRB05}) as a function of temperature.  Linear fit
of   temperature-dependent  resistivity   data   indicate  electron-phonon
interaction limited carrier transport in these laminated metals. }
\label{Fig1}
\end{figure}  

A number of scientific reports exist on the electronic structure calculations \cite{ZhouPRB00, HugPRB02, HugPRB05, MoPRB12}, phase stability \cite{EmmerlichPRB07, MusicPRB07, DahlqvistPRB10}, elastic \cite{WangPRB04, RaymundoPRB11} and thermal properties \cite{TogoPRB10} as well as NMR \cite{LuePRB06} studies of these laminated metals. On the contrary, very few experimental reports \cite{BarsoumPRB00, HettingerPRB05, MauchampPRB13, FinklePRB03} is devoted to the transport properties on these metals. Most of the reports assume empirical two band model to explain the experimental observation qualitatively without going into the microscopic details of carrier transport. Though there exists some reports on {\it {ab-initio}} studies on electronic structures, no efforts has been made to connect the electronic structures (bandstructure, phonon spectrum, etc) to the mechanism of carrier transport in these systems. A qualitative approach is made recently \cite{MauchampPRB13} to explain the anisotropic conductivity in Ti$_{2}$AlC, albeit without calculating electron-phonon interaction (assumes a constant scattering time) and phonon spectrum. Here, under deformation potential approximation and Boltzmann transport equation within relaxation time approximation, we develop an analytical expression of conductivity/mobility by carefully analyzing and extracting material parameters from  the electronic and phonon spectrum of MAX phase compounds (Ti$_{2}$AlC). Using our transport model, the extracted deformation potential values agrees well with the first principle calculations. The developed transport model as well as deformation potential value will be useful for predicting transport properties at elevated temperatures as well as optimizing the performances of devices made out of these laminated metals. Due to available of experimental data, we focus only on transport in Ti$_{2}$AlC as an illustrative example of MAX phase compound. Similar procedure can be used to  explain the measured transport data from electronic and phonon bandstructers of other MAX compounds as well as 2D Mxenes.

\section{Theory and Analysis}
\par
The  Fig.\ref{Fig1}(a) shows  the carrier  resistivity as  a  function of
temperature for  two different Ti$_{2}$AlC  metal films. It should be noted that
the carrier resistivity is increasing  with temperature - a typical signature
of metallic behavior.
\begin{figure}[t]
\includegraphics[width=85 mm]{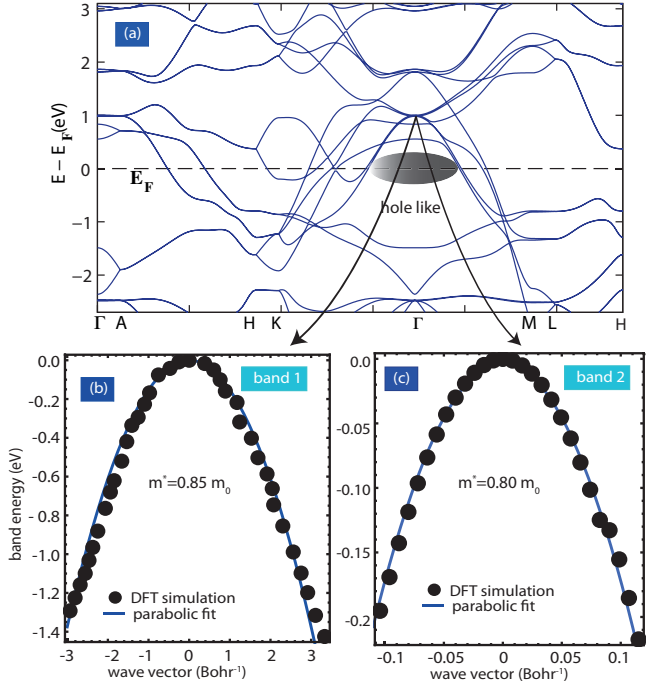}
\caption{(a)  electronic  bandstructure   of  Ti$_{2}$AlC  from  first
principle  calculations. Shaded area  shows the  Fermi level  and note
that holes  are the carrier  at the Fermi  level. Note that  along the
basal plane five  bands crosses the Fermi level  whereas along the $c$
axis ($\Gamma\rightarrow  A$) there exists  a insulating gap.  (b) and
(c) effective masses for cylindrical bands. }
\label{Fig2}
\end{figure} 
In Fig.\ref{Fig1}(b), we  plotted the inverse of mobility
(~ resistivity)  for higher temperatures.  The linear increase  of the
resistivity   with  temperature  indicates   electron-acoustic  phonon
scattering  limited  carrier transport  in  this  metallic system. These different set
of  measurements convincingly  indicates the  phonon-dominated carrier
transport  in  these  metal   films  as  resistivity  is  linear  with
temperature as  shown in Fig.1 (linear theoretical fit).  
 \begin{figure}[t]
\includegraphics[width=85 mm]{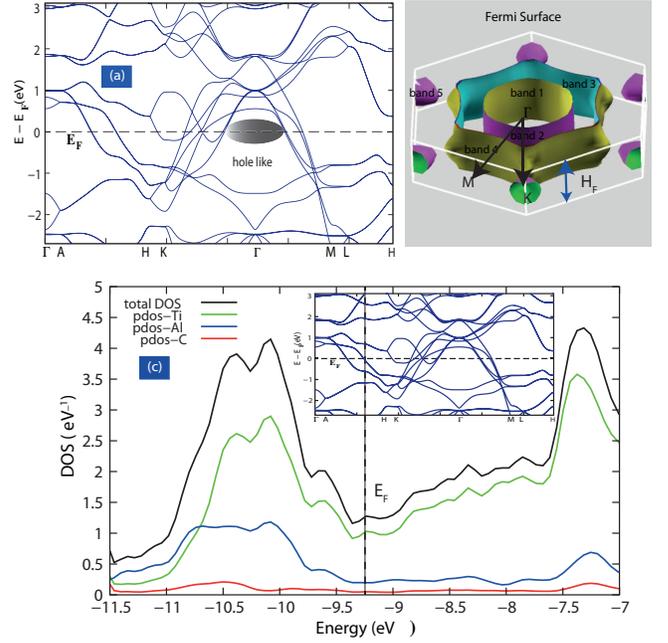}
\caption{(a) electronic bandstructure same as Fig.2 (a), and (b) Fermi
surface of  individual band  at the Fermi  level. Note that  there are
five  bands at  the Fermi  level and  each of  the surface  (inner and
outer) corresponds to one band. Iner two cylindrical bands responsible
of carrier  transport in this metal  along the basal  plane. (c) Total
density of  states (DOS) as  a function of energy.  Contributions from
each constituent atoms (Ti, Al and C) to the total DOS is plotted with
energy.  Note that  most of  DOS at  the Fermi  level coming  from the
d-bands of Ti. }
\label{Fig3}
\end{figure} 
\subsection{Electronic Bandstructure}
To investigate the  transport in these layered metals, it
is  extremely  important  to  know  electronic as well as phonon  band
structure of  the material. We leverage {\it{ab-initio}} density functional theory (DFT) to determine electronic structure. Figure \ref{Fig2}(a)  shows the electronic
bandstructure  of  bulk  Ti$_{2}$AlC.
Density functional theory (DFT) simulations were performed
within the local density approximation (LDA), employing
Perdew - Zunger (PZ) parameterization of the exchange-correlation
functional \cite{pz81} as implemented in the local orbital
SIESTA code \cite{SIESTA}. The core and valence electron interactions
were treated through pseudopotentials for all the atoms, using
Troullier - Martin (TM) method \cite{tm91}. Variable cell relaxation 
was performed till the forces between the atoms
were less than 0.02 eV/\AA.
Calculations have been carried out using a polarized
double zeta (DZP) basis with an energy shift of less than 15 meV
(0.001 Rydberg). A mesh cutoff of 400 Rydberg
and sufficiently fine k-point grid with $18\times18\times9$ 
Monkhorst - Pack mesh was used to achieve convergence.
Ti$_2$AlC has a hexagonal crystal structure with lattice parameters:
$a = b = 3.0691$ \AA, $c = 13.7364$ \AA, $\alpha = \beta = 90^{\circ}$,
and $\gamma = 120^{\circ}$ . The fully relaxed crytal within our generated
LDA-TM pseudo potentials has lattice parameters: $a = b = 2.9943$ \AA, 
$c = 13.4519$ \AA, $\alpha = \beta = 90^{\circ}$, and $\gamma = 120^{\circ}$.
These are within 2 - 2.5 percent of the experimental values, which is
to be expected as LDA typically underestimates the lattice parameters \cite{HassPRB09}.

The anisotropic bandtructure of   this   layered  metal   is   quite   evident   from  the   Figure
\ref{Fig2}(a). We notice that the  system behaves as a metal along the
layer (from  $\Gamma \rightarrow$ K,  Fermi level crosses  five bands,
and from $\Gamma \rightarrow$ M, Fermi level crosses  four bands),
while  along  the $c$  axis  (perpendicular to  the  layer)  it is  an
insulator (notice that no band crosses the  Fermi level along the
$\Gamma  \rightarrow$ A direction; there  exists a  virtual  band gap
along this direction). Moreover, it is evident from the band structure
that the transport is hole like  along the layer. These hole bands are
coming  from the  $d$-orbital  of the  titanium  (Ti) atom  in the  unit
cell. Among  these bands  in the Fermi  level, typically holes  in the
fast-moving  bands (bands  with steep  curvature thus  lower effective
masses) conduct  current. For these bands  the cumulative conductivity
is                              written                             as
$\sigma_{t}=\sum_{i}\sigma_{i}=ne^{2}\left[\sum_{i}\tau_{i}/m_{i}^{\star}\right]$. 
Here $n$ is the volume carrier density, $e$ is the electronic  charge,
$\tau_{i}$  is  the hole  lifetime  and  $m_{i}^{\star}$  is the  hole
effective  mass in  the $i^{th}$  band. As  conductivity  is inversely
proportional to the effective  mass, carriers in the highest curvature
band  (lower  effective  mass)   carries  most  of  the  current.   In
Fig.\ref{Fig2}(b) \& (c), we have shown the effective mass of two most
highest  curvature  (lowest effective  mass)  bands with similar effective mass $m^{\star}\approx 0.8m_{0}$, where m$_{0}$ is the free electron mass. It should be note that,  the band-edge effective mass definition  is only valid for semiconductors,
where Fermi  level lies  close to the  conduction band edge.   In this
case,  since the  fastest two-bands  are radially  symmetric  (see the
Fermi surfaces in Figure 3(b)), effective-mass formalism should be valid.
\subsection{Transport Formalism}
Having  determined the bandstructure, we now  turn into transport
calculations in this layered metal.  For an applied electric field $E$
(say  along  $x$ direction),  the  current in  a  metal is determined using
Boltzmann  transport  equation.  Under relaxation  time  approximation
\cite{BookZiman},  the current  density  can be  written as  $j_{x}=-\sigma(T)E$
where  $\sigma\left(T\right)   $  is  the   carrier  conductivity.   %
$f^{0}\left({\mathcal{E}}\right)$   is  the   equilibrium  Fermi-Dirac
distribution function,  and ${\mathcal{E}}$  is the band  energy.  The
carrier          conductivity           has          the          form
$\sigma(T)=2g_{v}\left(\frac{e}{\hbar}\right)^{2}{\displaystyle
\oint_{FS}}\left[\left(\frac{\partial{\mathcal{{E}}}}{\partial
k_{x}}\right)^{2}_{{\mathcal{E}}_{f}}\frac{\tau\left({\mathcal{E_{f}}}\right)dS}{\left(2\pi\right)^{3}\left|\nabla
{\mathcal{E}}\right|}_{{\mathcal{E}}_{f}}\right]$,  where  the surface
integration is done on the Fermi surface (FS). Here, $e$ is the change
of   an   electron,   $\hbar$   is  the   reduced   Planck   constant,
${\mathcal{E}}_{k}$ is the band energy  of carriers and $g_{v}$ is the
band degeneracy factor. The carrier relaxation time $\tau(k)$ contains
the  information  of electron-phonon  scattering  in  this metal.  The
electron-phonon  relaxation  time can  be  evaluated following  Mott's
two-band model\cite{BookMott}. Here holes from the symmetric fast band scatter
to the  vacant states which are coming from  the d-orbitals of
the  Ti.  Under  the   relaxation  time  approximation,  the  momentum
relaxation   due   to   phonon    scattering   can   be   written   as
$\tau(k)^{-1}=\frac{2\pi}{\hbar}{\displaystyle\int_{k}}\frac{d{{\bf{\vec{k}}}}}{\left(2\pi\right)^{3}}\lvert{\mathcal{M}}_{k,k'}\rvert^{2}\left(1-\cos\theta\right)\delta\left({\mathcal{E}}_{k'}-{\mathcal{E}}_{k}\right)$,
where $k$  ($k'$) is  the initial (final)  wave vector  of scattering,
${\mathcal{E}}_{k}$  (${\mathcal{E}}_{k'}$)  is  the  initial  (final)
carrier energy  and $\theta$  is the angle  of scattering.  The matrix
element  of  scattering   is  given  as  ${\mathcal{M}}_{k,k'}=\langle
k'|V(r)|k\rangle$,  where  $\langle k'|$  ($|k\rangle$)  is the  final
(initial)  state   of  scattering  and  $V(r)$   is  the  perturbation
scattering potential. For  long-wavelength crystal vibration (acoustic
phonon),  the   perturbation  potential is  $V(r)=
-\left[\sqrt\frac{2\hbar}{\Omega\rho\omega^{\lambda}(q)}\right]{\vec{q}}\cdot{\vec{e}_{\lambda}}\Xi\left[e^{i(k\cdot{\bf{r}}-\omega^{\lambda}_{q}t)}+c.c\right]$,
where $\Omega$  is the unit cell  volume, $\rho$ is  the mass density,
$\omega(q)$ is the phonon frequency, $q$ is the phonon wave vector,
${\vec{e}_{\lambda}}$  is  the  polarization vector, and $\Xi$ is the acoustic deformation potential \cite{BardeenPR50}.  
Denoting carrier Bloch wave   function   $\langle
k|r\rangle=\frac{1}{\sqrt{\Omega}}{\mathcal{U}}_{k}(r)e^{i{\vec{k}}\cdot{\vec{r}}}$,
the electron-phonon scattering rate for longitudinal phonon is written
as \cite{note}
\begin{equation}
\tau\left({\mathcal{E}}_{f}\right)^{-1}=\left[\frac{\pi\Xi^{2}k_{B}T}{\hbar\rho v_{s}^{2}}\right]N_{d}\left({\mathcal{E}}_{f}\right),
\end{equation}  where $k_{B}$ is  the Boltzmann  constant, $T$  is the
lattice  temperature, and $v_{s}$  is  the phonon  velocity. The entity
$N_{d}\left({\mathcal{E}}_{f}\right)$ is  the available d-band density
of states (available band for scattering in two band model) which are mostly coming from the  d-orbital of  the Titanium atom.  With this  prescription of
calculating scattering  time, it is possible to  determine the carrier
conductivity for  fast moving bands with cylindrical  symmetry (band 1
and band 2 in Fig.\ref{Fig3}(b)).  For the band with cylindrical Fermi
surface,    the    energy     dispersion    can    be    written    as
${\mathcal{E}}_{f}=\frac{\hbar^{2}(k_{F}^{R})^{2}}{2m^{\star}}$,   for$
-H_{F}/2\le  k_{z}\le H_{F}/2$,  where $H_{F}$  is the  height  of the
Fermi     cylinder    in    reciprocal     space    as     shown    in
Fig.\ref{Fig3}(b). Defining  elementary surface on  the Fermi cylinder
as  $dS=k_{F}^{r}d\phi dk_{z}$,  where we  use  cylindrical coordinate
system  $(k_{F}^{r},\phi,k_{z})$ for the  Fermi cylinder  and defining
the                  Fermi                 velocity                 as
${\bf{v_{F}}}=\frac{1}{\hbar}\nabla_{k}{\mathcal{E}}(k)=\sqrt{2{\mathcal{E}}_{f}/m^{\star}}\hat{r}$,
the conductivity integral can be simplified to
\begin{eqnarray} 
\sigma &=&\left[\frac{g_{v}e^{2}\tau\left({\mathcal{E}}_{f}\right)v_{F}k_{F}^{R}H_{F}}{\left(2\pi\right)^{3}\hbar }\right]\int dk_{z}\int_{0}^{2\pi}\cos^{2}\phi d\phi \\ \nonumber
&=&\left(\frac{2e}{h}\right)^{2}g_{v}\tau\left({\mathcal{E}}_{f}\right)H_{F}{\mathcal{E}}_{f}=\underbrace{\left(\frac{8m^{\star}{\mathcal{E}}_{f}H_{F}}{h^{2}}\right)}_{n_{h}}e\overbrace{\left[\frac{e\tau\left({\mathcal{E}}_{f}\right)}{m^{\star}}\right]}^{\mu_{h}},
\end{eqnarray} 
\begin{figure}[t]
\includegraphics[width=88 mm]{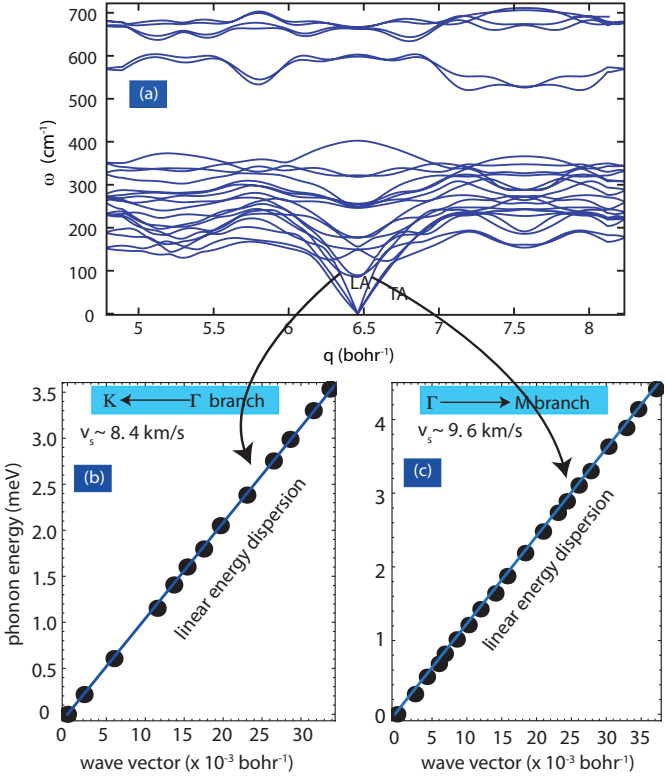}
\caption{(a) phonon bandstructure of Ti$_{2}$AlC calculated from first
principle  methods  illustrating   three  low-energy  acoustic  phonon
branches  (two  transverse  modes   and  one  longitudinal  mode)  and
dispersionless  optical modes. Longitudinal  acoustic modes  are fitted
with linear  dispersion relation ${\mathcal{E}}_{ph}=\hbar  v_{s}q$ to
extract sound velocity shown in (b) \& (c).  }
\label{Fig4}
\end{figure}    where    $n_{h}$    is    the   hole    density    and
$\mu_{h}=e\tau\left({\mathcal{E}}_{f}\right)/m^{\star}$  is  the  hole
mobility. Since there are two equivalent cylindrical bands, band degeneracy $gv_{v}=2$. Comparing  the  above   equation  with   the  conventional
definition of  conductivity $\sigma=ne\mu$, we write  the hole density
as   $n_{h}=    4m^{\star}{\mathcal{E}}_{f}H_{F}/h^{2}$.   Using   the
relations       $v_{F}=\sqrt{2{\mathcal{E}}_{f}/m^{\star}}$,       and
$k_{F}^{R}=\sqrt{2m^{\star}{\mathcal{E}}_{f}}$, we  have the following
form                  of                  hole                 density
$n_{h}=\left[(k_{F}^{R}/\pi)^{2}H_{F}\right]$.  This is  in sharp
difference  with isotropic  elemental  metals where  Fermi surface  is
rather   spherical    and   the   carrier   density    is   given   by
$n=k_{F}^{3}/3\pi^{2}$\cite{BookZiman}, where $k_{F}$ is the average isotropic Fermi wave vector. This stems  out from anisotropic bandstructure of these layered materials.
\par The  above two equations  pertaining to carrier transport  can be
compared with experimental observations to extract important transport
parameters. To  calculate the scattering  rate using Eq.1, we  need to
know few material related  parameters which are not a-priori available
in  literatures. For  example, the  scattering rate  calculation in
Eq.1  requires material  related  parameters such  as effective  mass,
Fermi  energy, d-band  density of  states at  the Fermi  level, phonon
velocity  and acoustic-phonon  deformation  potential energy.  Carrier
effective  mass  and  Fermi  energy  are obtained  from  the  ab-initio
bandstructure  calculation  as  shown  in Fig.\ref{Fig2}.  The  total d-band
density   of  states  $N_{d}\left({\mathcal{E}}_{f}\right)$   is  also
obtained  from  density functional  theory  calculations  as shown  in
Fig.{\ref{Fig3}(c).   Figure   \ref{Fig3}(c)   shows  the   individual
contributions  of  each constituent  atoms  to  the  total density  of
states. It  is evident from the plot that most of  the states in  the Fermi
level are coming from valence  d electrons of Titanium. Please note total density of states is plotted in Fig.4, whereas $N_{d}\left({\mathcal{E}}_{f}\right)$ is the density of states per unit volume. Using the unit cell volume $V_{c}=104.45$ Ang$^{3}$, the density of states at the Fermi energy (${\mathcal{E}}_{f}=-9.24$ eV) found to be $N_{d}\left({\mathcal{E}}_{f}\right)=1.19$x$10^{28}/eV.m^{3}$.
\subsection{Phonon Bandstructure} 
For the phonon velocity, we perform a  ab-initio phonon bandstructure calculations as
shown in Fig.\ref{Fig4}(a).  
To compute acoustic phonon modes, we employed the technique of frozen 
phonon method as implemented in SIESTA \cite{SIESTA}. Prior to this, we made sure the
convergence of all the ground state properties was achieved, and that the structure
was fully relaxed before we carried out phonon band structure calculation.
A large enough supercell with as many as 600 atoms was used to ensure 
sufficient attenuation of real-space force constants. 

Since the low-energy  acoustic phonon follow the linear
dispersion   rule,   i.e.   ${\mathcal{E}}_{ph}(q)=\hbar   v_{s}q$   (
${\mathcal{E}}_{ph}$ is the  phonon energy and $q$ is  the phonon wave
vector), by fitting the  longitudinal acoustic phonon (LA-phonon) with
a linear  function, we extract the average  phonon velocity $v_{s}\sim
9$  km/s.  The linear  fitting  of  LA-phonon  mode in  two  different
direction  is  shown in  Fig.\ref{Fig4}(b)  \&  (c).  Using all  these
calculated    parameters   values,    we   compare    the   calculated
temperature-dependent  mobility   with  experimental  measurements  to
extract  average  deformation   potential  value  for  electron-phonon
interaction.  Since  $\tau^{-1}\left({\mathcal{E}}_{f}\right)\sim  T$,
the inverse carrier mobility
$\mu^{-1}=\left[e\tau\left({\mathcal{E}}_{f}\right)/m^{\star}\right]^{-1}\sim
T$   as    shown   in   Fig.\ref{Fig1}.   By    fitting   the   linear
temperature-dependent   inverse  mobility  ($\sim$   resistivity),  we
extract the average acoustic-phonon deformation potential to be $\sim$
8 eV. Since  the scattering is interband scattering  (s-type to d-type
bands) in two bands models, this value
of  deformation  potential corresponds  to  the interband  deformation
potential  for electron-phonon  scattering.  The linear  fit of  inverse
mobility  (or resistivity) is  shown in  Fig.\ref{Fig1}(a) \&  (b). To
reconfirm this  value of  phonon deformation potential  extracted from
transport  data,   we  perform  an   density  functional  theory-based
computation    of    phonon     deformation    potential    in    this
material. 
We quantify the deformation potential \cite{cardona,vandewalle} through mimicking the 
lattice deformation due to phonons by introducing unit strain along
one or more lattice parameters, and then tabulate the band structure
change. Typically, we look at relative change in the conduction band minimum and 
the valence band maximum (Fermi energy for a metal).
We introduced uniaxial, biaxial, and volume compressive, tensile
strain as well as torsional strain to mimic all possible effects.
By definition \cite{vandewalle, ZungerPRB99,FioriAPL11} $\delta E = D_A \delta V/V$, where $D_A$ is the deformation
potential, $\delta E$ is the change in the band structure due to applied
strain (quantified through DFT), and $\delta V/V$ is the relative change in the
volume due to strain.
The  calculated values of  deformation potential $\sim$
6-8  eV are  in good  agreement with  the value  we estimated  from by
fitting the experimental data.
\section{Conclusion}
 In  conclusions, we have  investigated the microscopic  origin of
carrier  transport in  layered metals such as Ti$_{2}$AlC.  Density functional
theory  is  used  to  determine  the  electronic  as  well  as  phonon
bandstructure  of this  metal.  The electronic  bandstructure shows  a
anisotropic  behavior of this  material -  metallic conduction  in the
basal  plane and  an  insulating gap  along  the $c$  axis. Using  the
electronic and  phonon bandstructure calculated  from first principle,
we  formulate  the  electron-acoustic  phonon scattering  as  well  as
carrier  conductivity using  Mott's two  band approach.  Comparing our
transport equations  with temperature-dependent mobility measurements,
we estimate the interband  acoustic phonon deformation potential to be
around 8 eV.  This numbers agrees well with  the first principle-based
estimation.  Though we chose Ti$_{2}$AlC as a reference system in this work, using our methodology carrier dynamics and transport parameters can be calculated/predicted for a whole range \cite{HettingerPRB05} of MAX -phase compounds and two-dimensional MXene metals. Our transport formulations  and the  estimated parameters
will  be  useful  in  high-temperature application  of  these  layered
metals.


\begin{thebibliography}{}
\bibitem{note} The envelope wavefunction ${\mathcal{U}}_{k}(r)$ follows the normalization condition $\int_{\Omega}{\mathcal{U}}_{k}(r){\mathcal{U}}_{k}(r)^{\star}d{\vec{r}}=1$. For long wavelength acoustic phonon, phonon wavelength is much larger than the unit cell dimension. As a result, any integral like $\int_{V}{\mathcal{U}}_{k}(r){\mathcal{U}}_{k}^{\star}(r)f_{k}(r)d{\vec{r}}$, where $f_{k}(r)$ is a slowly varying function (such as long wavelength vibration) can be approximated as $\int_{V}f_{k}(r)d^{3}r\int_{\Omega}{\mathcal{U}}_{k}(r){\mathcal{U}}_{k}^{\star}(r)d{\vec{r}}=\int_{V}f_{k}(r)d^{3}r$. Note that here $V$ is crystal volume and $\Omega$ is the unit cell volume.
\end{thebibliography}
\end{document}